\def\year{2018}\relax
\documentclass[letterpaper]{article} %DO NOT CHANGE THIS
\usepackage{aaai18}  %Required
\usepackage{times}  %Required
\usepackage{helvet}  %Required
\usepackage{courier}  %Required
\usepackage{url}  %Required
\usepackage{graphicx}  %Required
\usepackage{color}
\usepackage{xspace}
\usepackage{float}
\usepackage{comment}
\usepackage{algorithmic}
\usepackage{amsmath,amssymb,amsthm,amsfonts}
\usepackage{times} 
\usepackage{enumitem}
\usepackage{bm}
\usepackage{xcolor}
\usepackage{tikz}
\usepackage{booktabs,fixltx2e}
\usepackage[flushleft]{threeparttable}
\usepackage{tabularx}

\makeatletter
\newcommand*{\rom}[1]{\expandafter\@slowromancap\romannumeral #1@}
\makeatother

\newenvironment{alg}{
    \begin{list}{}{
        \setlength{\itemsep}{2pt}
        \setlength{\parsep}{0pt}
        \setlength{\parskip}{0pt}
        \setlength{\topsep}{1pt}
    }
}
{
    \end{list}
}

\newcommand{\mc}[1]{\mathcal{#1}}

\floatstyle{ruled}
\newfloat{algorithm}{ht}{lop}
\floatname{algorithm}{Algorithm}
\makeatletter
\begin{document}
\title{Usability of Humanly Computable Passwords}
\author{
\begin{minipage}{4.6cm}
\begin{center}
Samira Samadi\\
\normalfont{Georgia Tech}\\
\normalfont{Atlanta, GA}\\
\normalfont{ssamadi6@gatech.edu}
\end{center}
\end{minipage}
\hspace{1cm}
\begin{minipage}{4.6cm}
\begin{center}
Santosh Vempala\\
\normalfont{Georgia Tech}\\
\normalfont{Atlanta, GA}\\
\normalfont{vempala@cc.gatech.edu}
\end{center}
\end{minipage}
\hspace{1cm}
\begin{minipage}{4.6cm}
\begin{center}
Adam Tauman Kalai\\
\normalfont{Microsoft Research}\\
\normalfont{Cambridge, MA}\\
\normalfont{adum@microsoft.com}
\end{center}
\end{minipage}
}

\maketitle
\begin{abstract}
Reusing passwords across multiple websites is a common practice that compromises security. Recently, \citeauthor{BlumV15} have proposed password strategies to help people calculate, in their heads, passwords for different sites without dependence on third-party tools or external devices. Thus far, the security and efficiency of these ``mental algorithms'' has been analyzed only theoretically. But are such methods usable? We present the first usability study of humanly computable password strategies, involving a learning phase (to learn a password strategy), then a rehearsal phase (to login to a few websites), and multiple follow-up tests. In our user study, with training, participants were able to calculate a deterministic eight-character password for an arbitrary new website in under 20 seconds.
\end{abstract}

\maketitle

\section{Introduction}

For over fifty years, passwords have served as the most common method of human-computer authentication and are likely to do so for the foreseeable future \cite{bonneau2015passwords}. Extensive research shows that many passwords in use can be easily guessed \cite{mazurek2013measuring} and that people reuse passwords across different accounts \cite{bonneau2012science}. Password reuse, though rampant in practice \cite{das2014tangled}, leaves accounts vulnerable to a single breach,  e.g., a malware attack to an unsecured website can lead to attacks to more important accounts if they enable an attacker to guess usernames and passwords.
In order to generate secure passwords, users have to create and remember complex strings, which often results in forgetting their passwords after a certain period \cite{weiss2008passshapes}. What makes this process even more tedious is that users~are often forced to change their passwords. Unfortunately, the number of unique and secure passwords that users can comfortably memorize is very limited \cite{florencio2007large}. To overcome this limitation, most users tend to choose simpler passwords, or one strong password and use it across multiple websites. These approaches have resulted in many password breaches over the past few years \cite{equifaxBreach,ebayBreach,adobeBreach,linkedinBreach,zappos}. 

In an attempt to ameliorate these difficulties, recent work has introduced mental password management schemas~\cite{BlockiKCD14,BlockiBD13,BlumV15,blocki2014towards} that enable users to systematically and securely generate and remember passwords for their different accounts. These schemas model passwords as mathematical functions from {\em challenges} (e.g., website names) to {\em responses} (character string passwords), and design such functions that can be computed by humans. Some of these schemas require paper or digital assistance \cite{blocki2014towards}, but we focus on those schemas that can be computed in one's mind without any additional resources.   

For brevity, we use the term \textit{mindhash} to refer to any such password management schema that enables a user to mentally compute a different password for each challenge without external memory or computational aid, i.e., without paper or a smartphone \cite{BlumV15}. Mindhashes require learning, memorization of a secret key, and execution when logging in to an account. In return for this effort, users enjoy security in the form of provable resilience to a small number of breaches.  \citeauthor{BlumV15} introduced several simple mindhashes with varying complexity, memory, and execution requirements, accompanied by varying security guarantees. We evaluate two of these mindhashes, one of which requires memorizing only three words. The schemas are resilient to multiple breaches in the sense that even knowing multiple different challenge-password pairs, an adversary is unlikely to be able to guess one's password to a different challenge. Moreover, these schemas are \emph{self-rehearsing} \cite{BlockiBD13} in the sense that the process of typing passwords on different websites naturally reinforces the user's memory of the secret key. 

Mindhashes may appear to be an appealing solution to the problem of remembering different passwords. Whether such methods are truly usable for most humans is an intriguing open question. Would human users (beyond mathematicians) find these methods pleasant? Would they be willing to adopt them? We address these questions through training and usability studies which are designed to teach mindhashes and measure their effectiveness. The amount of human computation required in executing these schemas was analyzed in precise models of human mental effort \cite{BlumV15}, but these formal models have yet to be tested in human experiments.

Several factors are important in the usability of such a system, including the amount of time that is required for learning and practice; memorizing the secret key; and using the mindhash to generate a password. It was suggested by \citeauthor{BlumV15} that a password strategy is humanly usable if ``any initial long-term memorization should take at most 1 hour, preferably less than 20 minutes; future rehearsals should take at most a total of 1 hour over the user's lifetime. Generation of a 10-character password should take at most 30 seconds, preferably less than 20 seconds.'' In~this work, we compare the usability of two mindhashes: a \emph{random-letter hash} (a simplification of other schemas from \citeauthor{BlumV15}) and a {\em 3-word hash} (called LP2 in \citeauthor{BlumV15}).
%We find that both mindhashes can be taught in a way that meets both desiderata. 
Following \citeauthor{BlumV15}, we also define \emph{security} of a password strategy by: (i) given no prior information, how difficult it would be for an adversary to guess any generated password, and (ii) given that an Internet hacker has access to a few passwords that are generated using a specific password strategy, how difficult it would be to guess a new password generated with the same password strategy. 

In our empirical user study, we teach participants how to use a mindhash using videos (less than 5 minutes) that explain the concept and the problem being solved and teaches them how to compute the mindhash function in general. We then help them to choose their secret key and to memorize it and have them practice using the mindhash on artificial website names. We later performed follow-up experiments simulating logins over the next month to evaluate how quickly and accurately participants can use their mindhashes on these and further artificial website names. 

We find that for the random-letter hash (3-word hash), the teaching phase involved  5.2 (4) minutes of videos,
a median of 8 (4.7) minutes to choose and memorize a secret key, a median of  6 (7.8) minutes to practice the mindhash on 15 logins. On these and 24 other logins performed over the next month, the median time to enter a password was 2.9 (3.2) seconds per character, and the mean success rate of typing the correct password within the first three tries was 98\% (91\%). Hence, there seems to be a tradeoff between learning and execution, with the 3-word hash being faster to learn and memorize a secret key, while the random-letter hash was faster to execute and gives higher accuracy.

The target audience of mindhashes is, potentially, anyone who seeks a secure way to remember different passwords across many different accounts. The participants in our usability study were US-based crowd workers on Amazon's Mechanical Turk crowdsourcing platform, which has been shown to source a diverse set of users \cite{stewart2015average,buhrmester2011amazon} and often produce results similar to those of more traditional approaches \cite{bentley2017comparing}. Nonetheless, such users have a certain minimum age and demonstrated the ability to learn to perform tasks (we filtered for 98\% task approval rating), which may differ from other groups of people using  multiple accounts. More specifically, our participants reported being between 21 and 55 years old, with the gender distribution of 40\% female and 60\% male. 

One experimental challenge is identifying whether (and how often) participants consult written or digital records of their secret keys. It is known that some people record passwords, and self-reporting cannot be relied upon, especially among Mechanical Turk workers \cite{peer2017beyond} who have concerns about bonuses and future work. To reduce the utilization of written records, we provided participants with a button that would, with a single press, remind them of their secret key while logging in. While this limits the ecological validity of our study, it enables us to track the frequency with which participants consulted this reminder. Participants varied in the frequency with which they pressed the secret key reminder button, though the frequency generally decreased (except among those participants that never pressed the button). The question of if users would keep records (and for what duration) is left for future work.

Mindhashes may be a viable alternative to the common password management approaches of password reuse or writing passwords down. Another solution to the password memorization problem is  to use a third-party password management software. Password vaults have become popular over the past few years as they require the user to remember only one master password and then the system automatically fills in login pages with strong (randomly generated) passwords. Unfortunately,~this~results in a single point of failure, which has caused security breaches \cite{gasti2012security}. Popular password vaults have been vulnerable to security attacks in recent years \cite{lastpass,oneloginbreach}. Moreover, the user must install the vault~on every~device that she uses, making it~difficult to~use on shared devices such as a~library computer~or~a~friend's phone. 

The rest of this paper is organized as follows. In Section~\ref{sec:passstrategies}, we define the random-letter and 3-word hashes. In Section~\ref{sec:design}, we describe the usability study in detail including the precise instructions given to the participants Then we present the results of the user study in Section~\ref{sec:results}. In Section~\ref{sec:theory}, we recall the human computation model of \citeauthor{BlumV15} and use it to analyze the usability and security of the random-letter and 3-word hashes. We discuss limitations of our study and mental password management schemas in Section~\ref{sec:limitations} and present conclusions and future work in Section~\ref{sec:conclusion}. 
%%%%%%%%%%%%%%%%%%%%%%%%%%%%%%%%%%%%%%%%%%%%%%%%%%%%%%%%%%%%%%%%%%%%%%%%%%%%%%%%%%%%%%%%%%%%%%%%%%%%%%%%%%%%%%%%%%%%%%%%

\section{Mindhashes}\label{sec:passstrategies}

Here we describe two mindhash functions and approaches to choose and memorize their secret keys. We will use these mindhashes in our empirical and theoretical analysis. Both mindhashes consist of a map from letters to letters. To generate a password, this character map is applied to the challenge (website name) left-to-right, and a special character string is appended that meets various password-composition policies. For example, if the website name is six characters and the special string is three characters, then the password will be nine characters, consisting of the application of the character map to each of the six characters of the website name followed by the three-character special string. \citeauthor{BlumV15} give more sophisticated mindhashes that have stronger security guarantees, but for the purposes of this study we restrict our attention to this character map type that still offers significantly higher security than reusing a small number of passwords. Note that for actual use, small modifications are necessary for special cases such as non-alphabetical characters or very short domain names, as discussed in Section \ref{sec:limitations}.

%%%%%%%%%%%%%%%%%%%%%%%%%%%%%%%%%%%%%%%%%%%%%%%%%%%%%%%%%%%%%%%%%%%%%%%%%%%%%%%%%%%%%%%%%%%%%%%%%%%%%%%%%%%%%%%%%%%%%%%%%%%%%%%%%%%%%%%%%%%%%%%%%%%%%%%%%%%%%%%%%%%%%%%%%%%%%%%%%%%%%%%%%%%%%%%%%%%%%%%%%%%%
\paragraph{3-word hash.}
For this mindhash, the secret key consists of a user-selected 3 words that in total contain at least 15 different letters of the alphabet, a random letter (which we will refer to as a wild card), and a special character string consisting of an uppercase letter, a digit, and a non-alphanumeric character. The three words are concatenated to one single string, called the 3-word string. For example, one secret key is shown in Table~\ref{table:3words}.
\begin{table}[h!]
\centering
\label{table:3words}
\begin{tabular}{| c | c | c |  }
\hline
  3-word string & wild card & special string \\
 \hline
  {\it adjust flight computer} & {\it x}  & {\it B7!}  \\
 \hline  
\end{tabular}
\end{table}	 
In this example, the 3-word string contains the 17 distinct letters 
$a~c~d~e~f~g~h~i~j~l~m~o~p~r~s~t~u.$
The character map takes any letter \emph{l} of the alphabet to the consonant that appears after the first occurrence of \emph{l} in the 3-word string. In case that the letter \emph{l} is not present in the 3-word string, then it maps it to the wild card. If the letter is the last consonant of the 3-word string, then it wraps around to the first consonant. Consonants are chosen because they offer greater entropy and hence greater security than vowels, which are more common and hence easier to guess.   

\begin{figure}[ht]
\centering
  \includegraphics[width=0.9\linewidth]{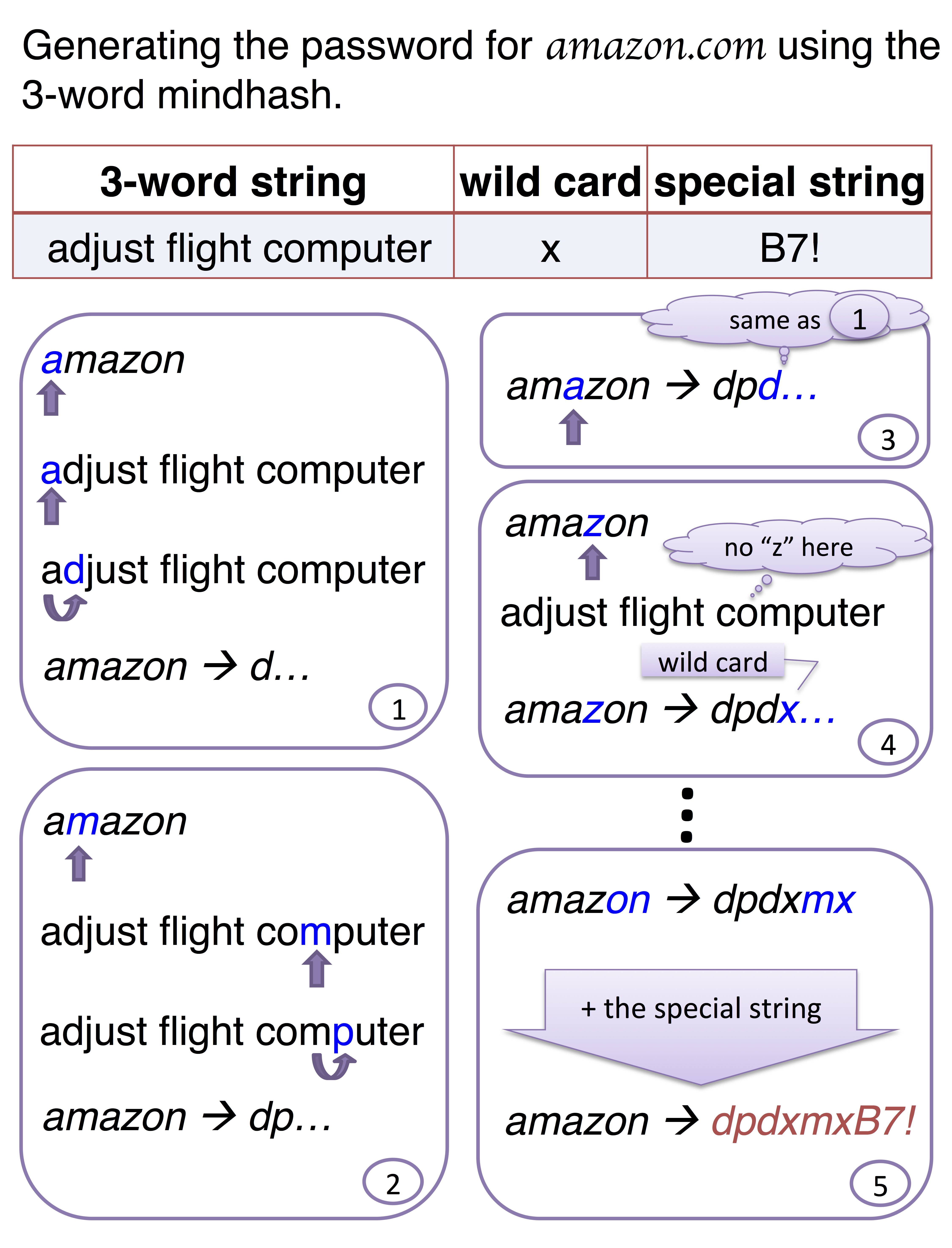}
  \caption{An example of generating a password using the 3-word mindhash. The top table shows the secret key and boxes 1-5 show the password generation step by step.}
  \label{fig:3wordImage}
\end{figure}

 Figure~\ref{fig:3wordImage} shows an example of how to apply the 3-word mindhash. Suppose that you want to login to {\it amazon.com}. The challenge is the word {\em amazon}.
\begin{itemize}
	\item Start with {\it a} (first letter of {\it amazon}) and find the first occurrence of {\it a} in {\it adjust flight computer}. Output the consonant that appears after {\it a}, which is {\it d} (Figure~\ref{fig:3wordImage}, box 1).
	\item The next letter is {\it m} and it appears in {\it computer}. The consonant after it is {\it p}. Output {\it p} (Figure~\ref{fig:3wordImage}, box 2).
	\item Repeat on the remaining letters of {\it amazon} (Figure~\ref{fig:3wordImage}, boxes 3-4).
	% (Figure~\ref{fig:3wordImage}, boxes 3-4). 
    \item Append the special string {\em B7!} (Figure~\ref{fig:3wordImage}, box 5).
\end{itemize}
Here are some examples of websites and their passwords.
\begin{table}[h!]
\centering
\begin{tabular}{| c | c | c  | }
\hline
  challenge & 3-word & Random-letter \\
 \hline
  {\it amazon} & dpdxmxB7! & qjqdr8*A \\
 \hline
  {\it facebook} & ldmrxmmxB7! & bqhgfddn8*A\\
 \hline
  {\it fidelity} & lgjrggfxB7!  & blcgply8*A\\
 \hline  
\end{tabular}
\caption{Passwords generated using 
our mindhashes.}
\label{table:passwords}
\end{table}	 

\paragraph{Random-letter hash.}
\label{sec:lettercodestrategy}
Like the 3-word hash, the random-letter hash is defined by a letter-to-consonant map and a special character string. Concretely, the user is aided in picking a \textbf{random} letter-to-consonant map for the first 20 letters of the alphabet (this truncation was done to increase usability -- considering the fact that frequency of the letters {\it uvwxyz} is about $8\%$ in total) and choosing a special 3-character string that meets common password-composition policy requirements. If the challenge contains a letter from {\em uvwxyz}, the user skips that letter without any output. Alternatively, one could map each of these letters to a wild card, but since these letters are not common, this is not necessary and is not considered here. For example, consider the map
\setlength\tabcolsep{3pt}
\begin{table}[h!]
\centering
\begin{tabular}{| c | c | c |  c | c | c |  c | c | c |  c | c | c |  c | c | c |  c | c | c |  c | c |}
\hline
  a & b & c & d & e & f & g & h & i & j & k & l & m & n & o & p & q & r & s & t  \\
 \hline
  q & f & h & c & g & b & s & k & l & m & n & p & j & r & d & t & n & w & x & y \\
 \hline  
\end{tabular}
\end{table}	

\noindent and the special string {\it 8*A}. Suppose that you want to login to {\it amazon.com}. The challenge is the string {\it amazon}. 
\begin{itemize}
\item Start with the first letter of the challenge {\em a} and find its mapping in the above table, {\em q}. Output {\em q}.
\item The next letter is {\em m}, output {\em j}.
\item Repeat on the remaining letters of {\it amazon}.
\item Append the special string {\em 8*A}.
\end{itemize}
See Table~\ref{table:passwords} for a few examples.
%%%%%%%%%%%%%%%%%%%%%%%%%%%%%%%%%%%%%%%%%%%%%%%%%%%%%%%%%%%%%%%%%%%%%%%%%%%%%%%%%%%%%%%%%%%%%%%%%%%%%%%%%%%%%%%%%%%%%%%%%%%%%%%%%%%%%%%%

\paragraph{Memorizing the 3-word hash.} The user memorizes the string of three words (order of words matters), a wild card and a special string. 

\paragraph{Memorizing the random-letter hash.} The user memorizes the letter hash using our method {\it Memorization with help of words}. The idea of this method is the following. The user looks at each letter pair, e.g., (a, q), and types the first word that comes to her mind that starts with the first letter and has the target letter as the next consonant, e.g., aqua. She will do the same for all letter pairs (Table~\ref{table:wordMemo1} left).

\begin{table}[!htb]
    \begin{minipage}{.5\linewidth}
      \centering
        \begin{tabular}{| l | l | l |  }
		\hline
  		a & q & aqua \\
 		\hline
 		b & f  & beef  \\
 		\hline  
 		c & h & chef \\
 		\hline
 		d  & c & duck \\ 
 		\hline 
 		e  & g & \ldots \\ 
 		\hline  
		\end{tabular}
    \end{minipage}%
   	\begin{minipage}{.5\linewidth}
      \centering
        \begin{tabular}{|l|l|}
            \hline
  a & aqua \\
 \hline
 b  & beef  \\
 \hline  
 c  & chef \\
 \hline
 d   & duck \\ 
 \hline 
 e  & \ldots \\ 
 \hline  
        \end{tabular}
    \end{minipage} 
    \caption{Memorization with help of words.}
     \label{table:wordMemo1}
\end{table}
Note that the mnemonics do not, in fact, have to be English words, but can be any memorable strings. Once the words are written for all pairs, the user only needs to memorize the (first letter, word) associations (Table~\ref{table:wordMemo1} right).
This part should be rehearsed with repetition, i.e., rote memorization. Once the (letter, word) associations are memorized, the user can directly use them to recover the letter hash. Finally, the user memorizes the special string.

%%%%%%%%%%%%%%%%%%%%%%%%%%%%%%%%%%%%%%%%%%%%%%%%%%%%%%%%%%%%%%%%%%%%%%%%%%%%%%%%%%%%%%%%%%%%%%%%%%%%%%%%%%%%%%%%%%%%%%%%%%%%%%%%%%%%%%%%%%%%%%%%%%%%%%%%%%%%%%%%%%%%%%%%%%%%%%%%%%%%%%%%%%%%%%%%%%%%%%%%%%%%%%%%%%%%%%%%%%%%%%%%%%%%%%%%%%%%%%%%%%%%%%%%%%%%%%%%%%%%%%%%%%%%%%%%%%%%%%%%%%%%%%%%%%%%%%%%%%%%%%%%%%%%
\section{Human Usability Study Design}
\label{sec:design}

In this section, we describe the details of our usability study. Participants were randomly divided into two groups, with half of the participants being assigned to each mindhash. The reader can access and try all our surveys at the following link: {https://github.com/PasswordUsability/Surveys}

\paragraph{Qualification.}
The participants had to pass a qualification test to be able to participate in our study. The qualification included reading a paragraph, informing about password security and then describing the study, followed by a  few simple multiple choice questions. 
The qualification tested that participants were paying attention and understood the need for having different passwords for different websites. 

\paragraph{Learning the 3-word hash.}
We showed the participants a short video teaching them how to generate a password with a 3-word mindhash. After the video, to make sure they understood the idea, we asked them to generate passwords for one website using the same secret key that was used in the video tutorial. Participants were provided with the secret key, multiple attempts, and hints to aid in learning.
At the end of this phase, participants learned how to generate passwords using a 3-word mindhash. After this phase, we asked the participants to choose their own three words sequence, wild card letter, and special string. Participants were allowed to proceed only if their word sequence contained at least 15 different letters and their special string contained an uppercase letter, a number, and a special character. As participants typed their 3 words, an alphabet letter bar, with the used letters crossed out, and the number of used letters was shown. This was to simplify the process of choosing words.

\paragraph{Learning the random-letter hash.}
\label{sec:RLMemorizationExp}
We showed the participants a short video teaching them how to generate a password using a random-letter hash. After the video, we displayed the letter map and the special string used in the video and asked them to generate two passwords. At this point, the participant did not need to memorize a letter map or a special string, but had to practice generating passwords using such a map. Next, we provided participants with an interface to choose a random letter-to-consonant map for the first 20 letters of the alphabet. 

In the next step, we showed them a simple illustrative video explaining our {\em memorization with help of words} technique.
Then we ask the participants to repeat the letter pairs and the corresponding words for themselves. Although such a memorization might be done more quickly by speaking aloud, we asked the participants to type the letter pairs and words to ensure compliance. To further solidify memorization of the character map, we gave three further exercises:
\begin{itemize}
\item Showing the letter pairs and asking the participants to type the words.
\item Showing only the left letter and asking the participant to first type the word and then the right letter.
\item The same as second exercise, but showing the left letters in a different~order,~e.g.,~``b, d, c, e, a'' in Table~\ref{table:wordMemo1}.
\end{itemize}

\paragraph{Practice.}
Immediately after the learning phase, participants were presented with 15 artificial website names to try to log in, one at a time. For each website, they were asked to type the password using the mindhash that they had learned. Two hint buttons were provided. One showed text instructions on how to generate the password using the mindhash, and the other one displayed the participant's secret key.

Participants had three tries to type each password. If they failed in all tries, they were presented with the correct response.  

\paragraph{Feedback after learning.}
Participants were asked to give us their feedback on different aspects of the study. We asked them if the task was fun/boring, easy/hard and if the password generation became easier toward the end, %atk. We used seven point scale bipolar rating for these survey questions as they have been shown to provide a good representation of constructs such as ease of use and satisfaction \cite{krosnick1997designing,muller2014survey}. 
on a seven point bipolar rating scale. 
Finally, we asked the participants whether they would like to participate in our follow-ups.

\paragraph{Follow-up evaluations.}
After learning and practice (day 0), we performed six follow-up evaluations of the participants' ability to log in using their passwords, over a period of one month. The first follow-up was performed the next day (day 1), the second follow-up was again a day later (day 2), and the remaining four follow-ups were at day 4, day 8, day 16 and the final follow-up during days 32-35. The last follow-up was scheduled during a holiday period and thus we allowed the participants to fill it out anytime during a 3 day interval. At each follow-up, the participants were asked to generate passwords for 4 challenges. For each challenge, three attempts was given to type the password, and then the correct password was shown. 

Studies show that users manage on average 25 password-protected accounts \cite{florencio2007large}. Some of these accounts are used frequently (e.g., work account) and some are used occasionally. 
We consider 25 synthetic website names chosen as random common words: \{kite, pillow, atlantic, bundle, reverse , family, quebec, cough, subject, mug, spike, fishing, jumper, knob, chord, quiz, fixed, world, campaign, warm, navy, banquet, hazy, chef, twist\}. We assume that the first 15 names are frequent accounts and the last 10 are occasional or newly opened accounts. To reflect this, we asked the participant to type the passwords for all the frequent accounts at the end of the learning phase. The challenges in the follow-up evaluations were chosen with probability $75\%$ from the frequent accounts and with probability $25\%$ from the infrequent accounts, to reflect the use of passwords for both logging in to frequent accounts and infrequent or one-time accounts. 
 
The 1/2/4/8/16/32-day timing follows a doubling schedule \cite{Pimsleur67}, which has been shown to be an effective repetition spacing in the practice of learning \cite{WozniakG94}.
In addition to these sequential follow-ups, we ran a quantitative follow-up survey on day 4 of the study. In this survey, participants were asked to provide a self-recall of the secret key that they memorized. 
  
\paragraph{Hints and writing down passwords.}
Since the study was performed online, one concern is that our results would be tainted by participants writing down their secret keys (or storing them in a file) and consulting this record, i.e., cheat sheet,
without our knowledge in the experiment. Moreover, self-reporting cannot always be trusted in an environment like Mechanical Turk when users have motivations that keep them from being honest
%, as has been shown in various experiments 
\cite{peer2017beyond,suri2011honesty}. In our case, workers do not know our experimental protocol and may hope for bonuses for ``good'' work, not to mention pervasive fears of unfair rejections. Moreover, workers who want to be on the good side of the requester for future work may try to impress the requester with their good memory. For all these reasons, it would be tempting for a worker to record his secret key without admitting it to a requester, even if the requester claimed that there would be no consequence for reporting this. To address this, participants knew that throughout the study they had constant access to two hint buttons, one reminding them of the instructions and the other one reminding them of their secret key. Participants were told that there was no penalty or cost
% ss: do we need "(other than that of pressing the button)"?
to use these hints, and they pressed the hint buttons liberally.
% Indeed some participants pressed the hint button liberally for each login.

%ss added 
Although we understand that pressing a hint button on the screen is ecologically different than using a cheat sheet, we were hoping that capturing participants' usage of these hint buttons could give us some insight about the extent of users long-term dependence to the cheat sheets in real-life.
The use of a written record may not constitute a serious security problem \cite{cheswick2013rethinking}, and this argument is of course only stronger if the written record is only consulted during the first few days of learning the secret keys.

%%%%%%%%%%%%%%%%%%%%%%%%%%%%%%%%%%%%%%%%%%%%%%%%%%%%%%%%%%%%%%%%%%%%%%%%%%%%%%%%%%%%%%%%%%%%%%%%%%%%%%%%%%%%%%%%%%%%%%%%%%%%%%%%%%%%%%%%%%%%%%%%%%%%%%%%%%%%%%%%%%%%%%%%%%%%%%%%%%%%%%%%%%%%%%%%%%%%%%%%%%%%

\begin{figure*}[t]
\centering
  \includegraphics[width=0.47\linewidth, height = 4.5cm]{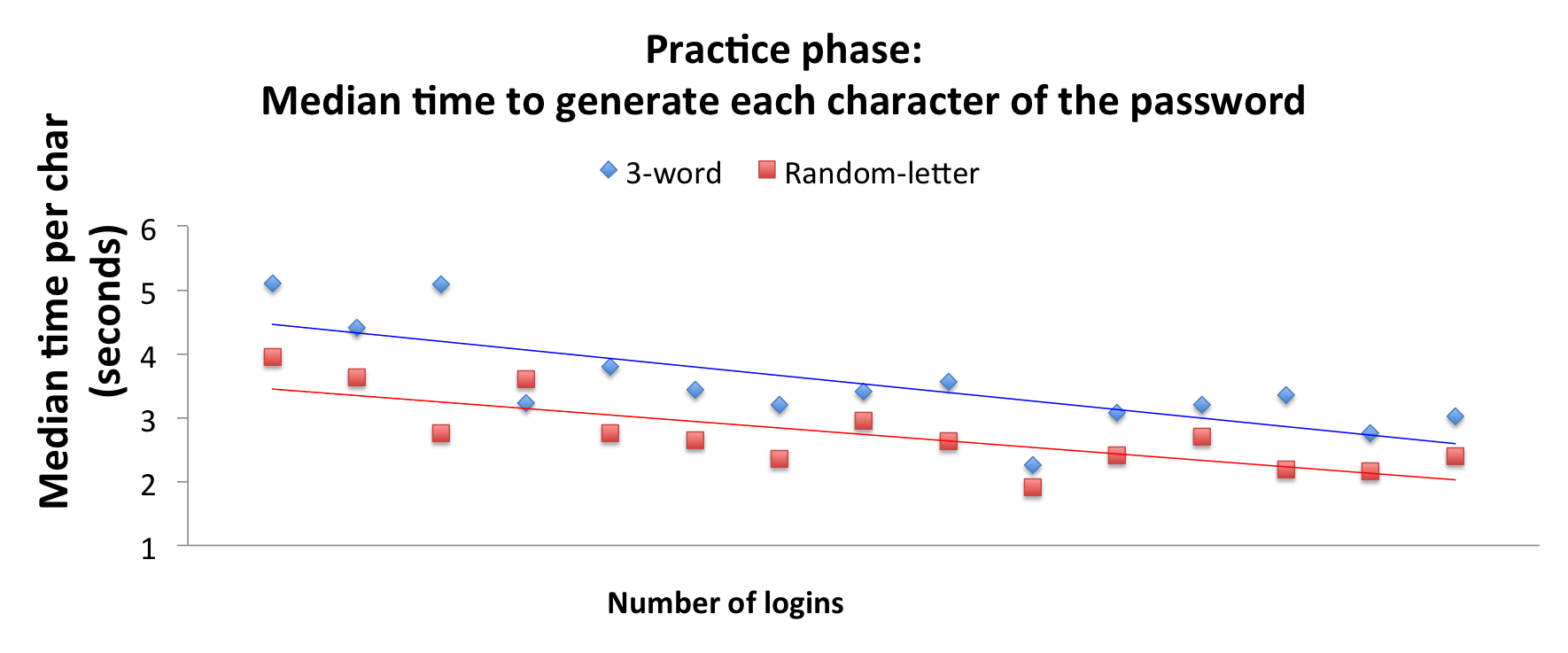}
  \hspace{.5cm}
  \includegraphics[width=0.47\linewidth, height = 4cm]{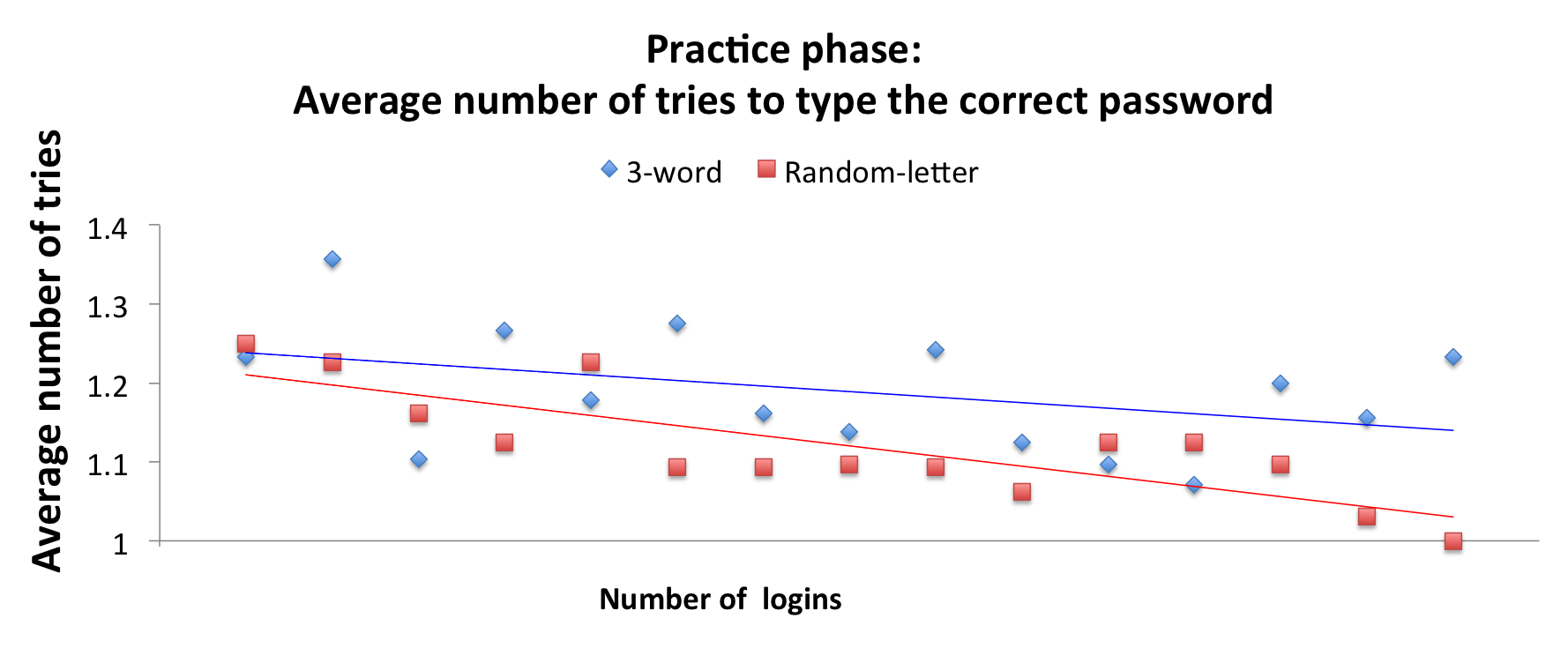}
  \caption{Practice phase, 15 logins. Left: Median time that the participants spent per character of the password. Right: Mean number of tries to type the correct password.}
  \label{fig:noTriesPractice_joint}
\end{figure*}

\section{Results}
\label{sec:results}
In this section, we present the result of our user study. The participants were US-based crowd workers on Amazon's Mechanical Turk platform with at least a 98\% task approval rating. Our participants reported being between 21 and 55 years old with average age of 31 years old. The gender distribution was 40\% female and 60\% male. For random-letter hash, overall 32 users participated in the training phase and 12 finished the last follow-up. For the 3-word hash, overall 34 users participated in the training and 14 finished the last follow-up \footnote{Participants were paid $\$2$ to complete the qualification, $\$10$ ($\$ 7$) to complete the day 0 training for random-letter (3-word) mindhash, $\$ 2$ for sequential follow-ups, and $ \$ 4$ for day 4 quantitative follow-up.}. Look at Table~\ref{table:participants} for more details.

\begin{table}[th]
\centering
\begin{tabular}{| l | c | c | c | c | c | c |c |   }
\hline
   mindhash/survey day & 0 & 1 & 2 & 4 & 8 & 16 & 32-35  \\
 \hline
  Random-letter & 32 & 27 & 27 & 24 & 14 & 14 & 12 \\
 \hline  
 3-word & 34 & 28 & 25 & 25 & 18 & 16 & 14 \\
 \hline
\end{tabular}
\caption{
Number of 
participants 
in the original study (day 0) and follow-up surveys during the one-month study. %atk one month of study. 
}
\label{table:participants}
\end{table}
\paragraph{Learning phase.}
\begin{table}[th]
\centering
\begin{tabular}{| l | c | c |  }
\hline
  Time & Random-letter & 3-word \\
 \hline
 Learning+Memorization & 8+13~min  & 11+0~min  \\
 \hline  
 Password~Generation & 19~sec & 25~sec \\
 \hline
\end{tabular}
\caption{
For random-letter (3-word) mindhash, learning time includes watching a 2 (4) minute tutorial video, choosing a personal secret key, and practicing the mindhash on a few passwords. Memorization time includes watching a 3.5 (0) minute video describing the memorization technique, and using it to memorize the secret key. Password Generation time is calculated for typing a password of length~8. 
}
\label{resultsTable}
\end{table}
Learning times are reported in Table~\ref{resultsTable}.  %atk It took the participants about 
The median times were 8 minutes to learn the random-letter hash and 11 minutes to learn the 3-word hash. The learning time for the 3-word hash was longer due to the longer training video (4-minute video versus 2-minute video). The memorization step for the 3-word hash was negligible. For the random-letter hash, the memorization time was 13 minutes, including a 3.5-minute video tutorial (see Section~\ref{sec:RLMemorizationExp} for the details of memorization).  

The median time that the participants spent on generating each character of the password decreases over time (Figure~\ref{fig:noTriesPractice_joint} left) with a mean of 2.3 seconds per character for the random-letter hash, and 3 seconds per character for the 3-word hash (computed over the last 5 logins). This corresponds to %atk implies 
a password generation time of 19 seconds for the random-letter hash and of 25 seconds for the 3-word hash for a password of length 8 (Table~\ref{resultsTable}) .

For both groups, the accuracy of typing the correct password during the practice phase was high: for each login, at least 96\% (82\%) of the participants typed correct passwords within three attempts using the random-letter (3-word) hash. Furthermore, the mean number of tries decreased over time (Figure~\ref{fig:noTriesPractice_joint} right).

\begin{figure*}[t]
\centering
  \includegraphics[width=0.46\linewidth, height = 4cm]{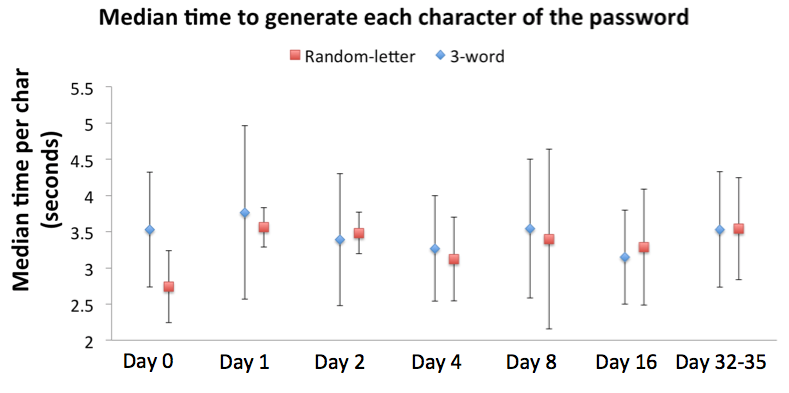}
  \hspace{1cm}
    \includegraphics[width=0.46\linewidth, height = 4cm]{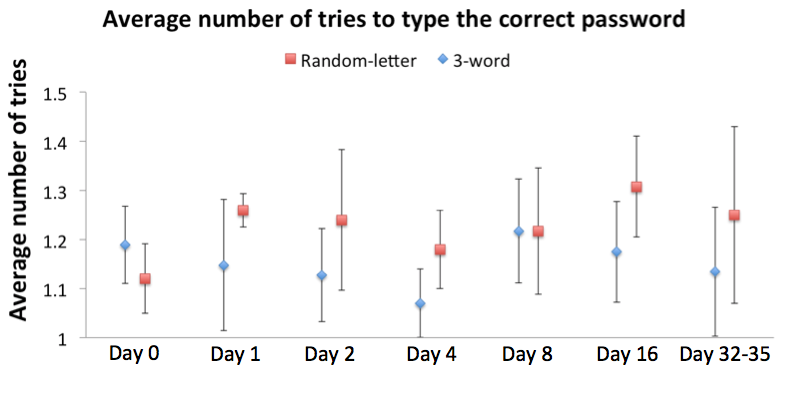}
  \caption{Left: Median time that participants spent generating each character of the password.  
Right: 
  Mean number of tries for participants to login successfully.
  Error bars represent one standard error.}
  \label{fig:noTriesErrorBar_joint}
\end{figure*}

In Figure~\ref{fig:noTriesPractice_joint} we see that, for both 3-word and random-letter mindhashes, the speed of generating each character of the password (left figure) and the accuracy of typing the password (right figure) increases as the number of logins increases. This improvement is an indicator that these mindhashes are self-rehearsing. 
\paragraph{Follow-ups.}
Figure~\ref{fig:noTriesErrorBar_joint} left shows the median of the time that the participants spent on generating each character of the password each day. The error bars indicate the standard deviation of the medians, across participants, for all logins during the day. From prior work on memorization and self-rehearsing passwords, we hypothesized that the passwords generation time would decrease over time as the secret key and password process is establishing in long-term memory. Indeed, for both mindhashes, although the gaps between follow-ups doubled each time, password generation time remained low (less than 3.5 second/character during the last follow-up). This is an evidence that users could still type passwords reasonably quickly even for websites that are visited rarely. Figure~\ref{fig:noTriesErrorBar_joint} right shows the mean number of tries that participants needed to successfully login. Our result shows that, although the gaps between logins  
doubled over time, the accuracy of typing the correct password for both mindhashes remained high over time (less than 1.3 attempts to successfully login). This is consistent with the self-rehearsable property of the password schemes.

\begin{figure}[ht]
\centering
  \includegraphics[width=0.9\linewidth, height = 4cm]{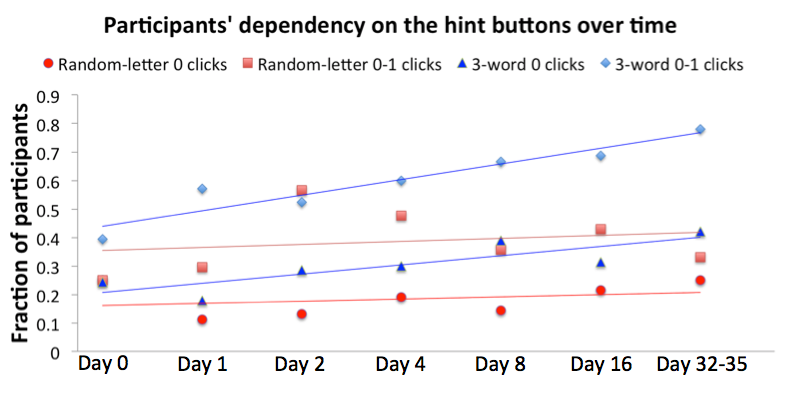}
  \caption{
  Fraction of participants that did not click on the hint button during logins and the fraction of participants that used the hint button for maximum one login.}
  \label{fig:0-1click}
\end{figure}

As the participants type their passwords over time (follow-ups), we expect the secret key to be self-rehearsed and therefore participants to click on the secret key hint button less frequently. Figure~\ref{fig:0-1click} shows the fraction of participants that used the secret key reminder hint button. 
Table~\ref{hintTable} shows these results for the last follow-up. For 3-word hash 42\% of the participants successfully typed passwords without the help of the hint buttons. For random-letter hash, the number is smaller, 25\%, but still comprises a meaningful fraction of users. 
Note that participants were told that there was no penalty for using the hints, hence the actual use of such aids in practice would be expected to be lower.
\begin{table}[ht]
\centering
\begin{tabular}{| l | c | c |  }
\hline
 No. click(s) & Random-letter hash & 3-word hash \\
 \hline
 0 clicks & 25\%  & 42\%  \\
 \hline  
 $\leq$ 1 click& 33\% & 78\% \\
 \hline
\end{tabular}
\caption{
Fraction of participants that clicked on the hint button for the secret key  during the last follow-up. 
}
\label{hintTable}
\end{table}
At the 4th day quantitative follow-up, participants were asked to type a free recall of their secret key. For the 3-word hash, $70\%$ of the participants perfectly remembered the 3 words and $95\%$ of the participants remembered at least 2 words. For random-letter hash $31\%$ of the participants remembered at least 18 letters out of 20 ($90\%$ of what they memorized), and $78\%$ of the participants remembered at least 12 out of 20 letters ($60\%$ of what they memorized). 

At the end of the follow-ups after a month, participants were asked if they had adopted the mindhash for generating  passwords for managing their own personal passwords. 
 For random-letter hash, 25\% of the participants reported  that they have adopted the mindhash in ``real life''. For the 3-word hash, 42\% of the participants reported that they used the mindhash for generating their own personal passwords. 
Although such statistics are known to be greatly inflated, the comparison between the two schemes may be of interest.

\paragraph{Feedback.}
At the end of the training phase, we asked the participants to fill out the feedback form discussed in Section~\ref{sec:design}, and we further received free-text feedback  
throughout the one month study. For both mindhashes, participants reported that the effort for generating  password  
decreased over time. Participants of both studies  reported that they have found the task of generating passwords using mindhashes \emph{neither easy nor hard}, with random- letter hash being slightly easier. Participants of random-letter hash found the task slightly fun. This was not completely the case for the 3-word hash as the participants reported that the task was neither boring nor fun. For both studies, 6\% of the participants reported that they wrote down information.

Overall, participants found the random-letter hash study more fun and interesting. One reason was that they were surprised that they could memorize such a letter map: ``This actually worked'' or ``Worked surprisingly well'' were some of the comments that they provided. Over the one month period, participants got comfortable generating passwords and reported that the password generation is feeling more and more natural over time. Some typical 
anecdotal feedback that participants provided 
during the follow-up included:\\
{\bf 3-word hash:} ``It's getting easier.'' or ``It's definitely getting easier.  I still have to open my word list but my brain is adapting and I'm starting to know what each letter should translate to without looking sometimes.''.\\ 
{\bf Random-letter hash:} ``I have definitely warmed up to the program.  It feels more natural now than the last time. '' or ``I think I've finally got a handle on this password combination! Well, minus the one mistake.''.

%%%%%%%%%%%%%%%%%%%%%%%%%%%%%%%%%%%%%%%%%%%%%%%%%%%%%%%%%%%%%%%%%%%%%%%%%%%%%%%%%%%%%%%%%%%%%%%%%%%%%%%%%%%%%%%%%%%%%%%%%%%%%%%%%%%%%%%%%%%%%%%%%%%%%%%%%%%%%%%%%%%%%%%%%%%%%%%%%%%%%%%%%%%%%%%%%%%%%%%%%%%%
\section{Theoretical analysis}
\label{sec:theory}
Usability of a mindhash has two main aspects: learning time and password generation time. In this section, we discuss the rigorous model from \citeauthor{BlumV15} for the password generation time.

\noindent\textbf{Password generation time} is the time that the user spends on outputting her passwords. Password generation is done entirely in the human's head with no paper, writing instrument, 
or computing device. It can be viewed as a restricted streaming computation. The working memory \cite{jonides2005processes} is very small, typically at most one or two pointers and two characters (which might typically be letters or digits). Each elementary operation (retrieve a sequence from long-term memory, follow a pointer, add two digits mod 10) has a cost, which is the total number of write operations to the working memory. For example, retrieving a pointer to a sequence in long-term memory has cost $1$, following the sequence has cost $1$, adding two digits $\pmod{10}$ has cost $1$ or $2$ depending on the number of digits created. A human algorithm can thus be assigned a total cost, by adding up the cost of each~step. This is the human complexity of the algorithm (called 
Human Usability Measure or HUM in \citeauthor{BlumV15}). It is meant~as a complexity measure for human computation analogous to the standard runtime complexity analysis of Turing machines. 
The HUM measures the human effort required to execute algorithms.
Just as machines running the same algorithm can take different times, humans also have variability in speed. 

\noindent \textbf{Password Generation Phase.}
Given a challenge $c=c_1\ldots c_n$, start with the first letter $c_1$. Output the mapping of $c_1$.
If mapping of $c_1$ is not defined, don't output anything or output the wild card (depending on the mindhash's instruction). 
Shift the pointer to the next letter and do similarly for the remaining letters. Append the special string $s$ to the end of your password.  
To illustrate this measure, we now compute the HUM for the 3-word and the random-letter hash.

\paragraph{HUM of 3-word hash.}
Let $W_1$$W_2$$W_3$ be the sequence of words and $s$ the special string. Let $f$ be the letter-to-letter map defined by the 3-word hash. 
% We can approximate the cost of applying the map $f$ to the letters of the challenge by ${l_w}/{2}$. 
The cost of applying $f$ is initially higher (to scan the words and find the next consonant) but finally becomes $1$.

\begin{algorithm}
\begin{alg}
\item[] Input: Challenge $c = c_1 \ldots c_n$
\item[] Retrieve challenge $c$. Pointer $\rightarrow c_1$. \hfill Cost = $1$
\item[] While not end of $c$ :
\begin{itemize}
    \item[] Let $c^*$ be the current character. 
	\item Output $f(c^*)$  \hfill Cost = $1$ 
	\item Shift pointer to next character. \hfill Cost = $1$
\end{itemize}
\item[] Retrieve fixed string $s$. Pointer $\rightarrow s_1$.  \hfill Cost = $1$
\item[] While not end of $s$ :
\begin{itemize}
	\item Output current character. \hfill Cost = $1$
    \item Shift pointer to next character. \hfill Cost = $1$
\end{itemize}
\end{alg}
\caption{3-word hash}\label{alg:HUM}\label{alg:3word}
\end{algorithm}
The HUM is $1 + n(1 + 1)+ 2|s| = 2n + 2|s| + 1$. 

\paragraph{HUM of random-letter hash.} 
Similar to Algorithm~\ref{alg:3word}, we can write the algorithm that describes the HUM of the random letter~hash. The HUM is $2n+2|s|+1$.
\begin{comment}
\begin{algorithm}
\begin{alg}
\item[] Input: Challenge $c = c_1 \ldots c_n$
\item[] Retrieve challenge $c$. Pointer $\rightarrow c_1$. \hfill Cost = $1$
\item[] While not end of $c$ : 
\begin{itemize}
    \item[] Let $c^*$ be the current character
	\item Output $f(c^*)$  \hfill Cost = $1$ 
	\item Shift pointer to next character. \hfill Cost = $1$
\end{itemize}
\item[] Retrieve fixed string $s$. Pointer $\rightarrow s_1$.  \hfill Cost = $1$
\item[] While not end of $s$ :
\begin{itemize}
	\item Output current character. \hfill Cost = $1$
    \item Shift pointer to next character. \hfill Cost = $1$
\end{itemize}
\end{alg}
\caption{Random-letter hash}\label{alg:HUMLC}
\end{algorithm}	
\end{comment}

\noindent \textbf{Security.}
Password strategies should be secure against a computationally all-powerful adversary observing (challenge, response) pairs and trying to impersonate the human. We use the following two security parameters, identified in earlier work \cite{BlumV15}. 
 
 It should be hard for the adversary to guess any password of the user. This is the intuition behind the definition of the security parameter $K$. Given a password strategy $\mc{S}$ and a positive integer $i$, we say that $K_{\mc{S}} = i$ if for any single challenge $c$, the probability that an adversary can guess the correct response to $c$ is at most $1/{10}^i$.
 	
 Assume that an internet hacker has found your password to a couple of insecure websites, and is trying to login to your bank account. She might not have the full information to precisely guess your bank account password, but she will have partial information that narrows her predictions to $4$ choices. As a result, if your bank website allows her to try multiple guesses, she can successfully login to your account. How many tries will she need? How many passwords will she need to see in the clear? This is the motivation behind the definition of the security parameter $Q$. Given a password strategy~$\mc{S}$ and $P\in (0,1)$, $Q_{\mc{S}}(P)$ is defined as the number of random (challenge, response) pairs that an adversary must observe in order to be able to respond correctly to the next challenge with probability greater than $P$. The dictionary of challenges must be specified (e.g., English words, random strings, the top 500 most popular website names, etc.).

%It is important to note that there is an inherent tradeoff between security and usability of any password strategy. Generating more secure passwords requires the user to memorize more information. 

\paragraph{{\bf $K_{\text{random-letter}}$}.} Given a challenge $c=c_1 \ldots c_n$, the adversary can respond correctly to $c$ only if she can correctly guess the mappings for all the letters of $c$ and the special string $s$. Each random letter has been chosen uniformly at random from the set of 21 consonants. The user's special string consists of one capital letter, one number and one special character, all chosen uniformly at random too\footnote{This assumption is based on the distribution of special strings reported by the users.}. Therefore, the probability that the adversary can guess the correct password is
\begin{align*} 
Pr[\text{guess}(c) =  \text{password}(c)] &\leq (1/21)^{n}(1/26)(1/10)^2.
\end{align*}
Assuming that an average password has length 8 (challenge of length five characters\footnote{For longer challenges, the user can use only the first 5 characters of the challenge.}), this gives us $K_{\text{random-letter}} \geq 10$.

\paragraph{{\bf $Q_{\text{random-letter}}(P)$}.} The adversary can respond correctly to a challenge only if she has seen all letters in the challenge in the previous challenges. If she has not seen even one letter, the chance of guessing the correct response to the challenge is $1/21$. What are the expected number of (challenge, response) pairs that the adversary should see to have complete knowledge of the mapping of all letters of a random new challenge?
% This value depends on the dictionary of challenges. 
For the top 500 domain names, this value is equal to $6.6$ \cite{BlumV15}. Therefore, for any $P \geq 1/21$,
$
Q_{\text{random-letter}}(P) \geq 6.6.
$  

Most secure websites block the user's account if he types a wrong password for 3-5 times. This is equivalent to $20\% <P < 33\%$, and thus the above security parameter value is meaningful. Also note that once the adversary sees one password, she already knows the special string $s$. Therefore $s$ does not contribute to $Q$. The security of the 3-word hash is lower since the total entropy generated by choosing 3 random words is smaller.
% \cite{BlumV15}.
%(estimated as between $3$ and $4$ in \citet{BlumV15}). 
$Q_{\text{3-word}}$ is estimated as between $3$ and $4$ \cite{BlumV15}.

%%%%%%%%%%%%%%%%%%%%%%%%%%%%%%%%%%%%%%%%%%%%%%%%%%%%%%%%%%%%%%%%%%%%%%%%%%%%%%%%%%%%%%%%%%%%%%%%%%%%%%%%%%%%%%%%%%%%%%%%%%%%%%%%%%%%%%%%%%%%%%%%%%%%%%%%%%%%%%%%%%%%%%%%%%%%%%%%%%%%%%%%%%%%%%%%%%%%%%%%%%%%%%%%%%%%%%%%%%

\section{Limitations}\label{sec:limitations}

We have shown that mindhashes are secure and human-usable solutions for choosing passwords for many users. However, in this section, we discuss limitations of mindhashes and the study that we have done in this paper. 

\noindent
{\bf Password policies.} Websites have policies with different password requirements involving password length or special characters. In our study, users were instructed to append a fixed ``special string'' to all of their passwords in order to meet such requirements. A recent survey finds that it is often possible to choose a single such string that simultaneously satisfies the requirements of different websites \cite{SeitzHPS2017}. However, in special cases, a website may have different requirements that may not be met by the special string. In general, this is considered as a challenging problem for other password generating approaches as well \cite{furnell2007assessment}.\\ 
{\bf Short or irregular challenges.} Some website names may be very short or contain non-alphabetic characters, such as 53.com for the \emph{Fifth Third Bank}. While not measured in our study, it would be natural for users to choose a memorable, sufficiently long challenge for these websites, such as the string {\em fifththird}. Note that different users may choose different challenges, but this does not cause any problems as long as each user is consistent with using the same challenge. Further study is necessary to see how common this problem is and how easy it is for users to recall their challenges.\\
% I'm not sure I would argue for writing things down, even though that is practical and safe, as that brings us into a different space of solutions  -- Adam
{\bf Infrequently used accounts.} Our study does evaluate the ability to correctly generate passwords for numerous new challenges, which is similar to generating a password for a rarely visited website. The 3-word hash has a natural self-rehearsing property so that using it frequently reinforces the memory of the entire secret key, and hence generating passwords for rarely used challenges is straightforward. However, for the random-letter hash, {\em infrequently used letters} pose a greater problem. For example, a user may forget her mapping for the letter {\em q} if it is never used.\\  
{\bf Passwords sharing.} Sharing passwords across different accounts is a problem that is not addressed by the mindhashes. Although mindhashes do not offer any solution for sharing passwords across different accounts, if a user chooses to share a password, security is not entirely compromised.\\
{\bf Changing passwords.} Certain systems may require passwords to be changed periodically. This is a problem with the password management that is not studied in our work and is not directly addressed by mindhashes. A solution for this, suggested in \citeauthor{BlumV15}, is to append a digit that indicates which letter of a challenge the user should start with when generating a password. The human usability of these approaches can be studied as part of  future work.\\  
% Again, I'm not sure we should argue for writing things down even though that is practical. For me, it would be easier to use an entirely different approach for remembering my work password which I have to change every few months    
{\bf Entropy of passwords.} Mindhashes assume that the secret key is chosen randomly. For example, in the random-letter strategy, we assume that the user memorizes a random letter-to-letter map. Although we provide a simple interface for users 
to build such a random map, it is still possible that in practice users may choose predictable secret keys (e.g., for the letter {\em a} some letters may be more commonly chosen, such as {\em p} for {\em apple}, or a person named {\em Alice}
may be more likely to choose {\em l}). This would reduce the entropy and %atk gives an 
advantage an adversary that attempts to guess the secret~key.\\ 
{\bf Multiple accounts on the same website.} Some users may have multiple accounts on one website. In this case, they may use the same password across accounts.\\
{\bf Dropouts and hints.} For both mindhashes, approximately 60\% of participants  dropped out during the course of the study. Our statistics should be interpreted as representative of the 40\% of participants who completed the study. 
While we could have provided additional incentives in the form of completion/milestone bonuses to increase completion rates, we felt that there was value in observing the natural completion rate at a static pay rate. As discussed,
participants had the opportunity to press a hint button to see their secret keys without any discouragement or adverse affect on their payment. In the last follow-up, 25\% (42\%) of the participants using the random-letter (3-word)
mindhash did not use hints even once. Taken together, if one considers mindhashes ``usable'' for such participants, this gives a {\bf lower bound} of 9\% (18\%) on the usability rate. This is a lower bound because it is likely 
that some users did not complete the study for various personal reasons aside from usability, and that some users clicked the hint buttons even when they would have found the system usable without hints.  
We provided
the hint button to dissuade users from secretly recording their secret keys in a way that we could not monitor.

%%%%%%%%%%%%%%%%%%%%%%%%%%%%%%%%%%%%%%%%%%%%%%%%%%%%%%%%%%%%%%%%%%%%%%%%%%%%%%%%%%%%%%%%%%%%%%%%%%%%%%%%%%%%%%%%%%%%%%%%%%%%%%%%%%%%%%%%%%%%%%%%%%%%%%%%%%%%%%%%%%%%%%%%%%%%%%%%%%%%%%%%%%%%%%%%%%%%%%%%%%%%%%%%%%%%%%%%%%
\section{Conclusion}
\label{sec:conclusion} 

This paper presents the first user study of two different mindhashes (i.e., password strategies): \emph{3-word hash} and \emph{random letter hash}. Participants in our user study spent a median of 11 minutes learning the 3-word hash and 8+13=21 minutes learning the random-letter hash. After the learning phase, the user is ready to use these mindhashes, and it takes 19-25 seconds to generate a password.
As predicted by the self-rehearsing property of mindhashes, the time to generate a password decreases over time. We showed that, although there are increasing gaps between rehearsals with no practice, users remembered their secret key and were able to successfully login to arbitrary websites. Although the presence of the reminder button decreases ecological validity, users consulted the button with decreasing frequency.  It was encouraging that some users seemed interested in adopting these methods to manage their own passwords. A natural research question is to identify mindhashes with even better usability and security. 

\paragraph{Acknowledgement.} This work was supported in part by Microsoft Research and NSF awards CCF-1563838 and CCF-1717349. We would like to thank Manuel Blum, Vivek Sarkar, and Rosa Arriaga for helpful discussions on the topics of this paper.

\bibliographystyle{aaai} 
\bibliography{passref}

\end{document}